\documentclass[11pt]{article}

\usepackage{latexsym}
\usepackage{amsmath}
\renewcommand{\epsilon}{\varepsilon}

\usepackage{amsmath,bm}
\usepackage{amssymb}
\usepackage{xcolor}
\usepackage{amsthm}
\usepackage[ansinew]{inputenc}
\usepackage{graphicx}
\usepackage{epsfig}
\usepackage{dsfont}
\usepackage{bbm}
\usepackage{url}
\usepackage{placeins}
\usepackage[FIGTOPCAP]{subfigure} 
\usepackage{multirow}

\usepackage{overpic}

\usepackage[authoryear]{natbib}
\usepackage[colorlinks]{hyperref}
\hypersetup{allcolors =blue}
    
\newtheorem{satz}{Theorem}[section]

\newtheorem{algorithm}[satz]{Algorithm}

\def\3{\ss}

\usepackage{dsfont}
\newcommand{\eins}{\mathds{1}}
\newcommand{\NN}{\mathds{N}}
\newcommand{\RR}{\mathds{R}}
\newcommand{\EE}{\mathds{E}}
\newcommand{\Prob}{\mathds{P}}

\newcommand{\bea}{\begin{eqnarray*}}
	\newcommand{\eea}{\end{eqnarray*}}
\newcommand{\be}{\begin{eqnarray}}
\newcommand{\ee}{\end{eqnarray}}

\newcommand{\ba}{\begin{array}}
	\newcommand{\ea}{\end{array}}

\def\3{\ss}

\begin{document}
	
	\title{{\bf \Large Testing for similarity of multivariate mixed outcomes using generalised joint regression models with application to efficacy-toxicity responses}}

\author{Niklas Hagemann$^{1,2}$,
        Giampiero Marra$^{3}$,
        Frank Bretz$^{4,5}$ and
        Kathrin M\"ollenhoff$^{2}$  \bigskip\\
	\small $^{1}$ Mathematical Institute, Heinrich Heine University D\"usseldorf, Germany \\
 \small{$^{2}$ Institute of Medical Statistics and Computational Biology (IMSB),}  \\ \small Faculty of Medicine, University of Cologne, Germany \\
    \small $^{3}$ Department of Statistical Science, University College London, UK \\
    \small $^{4}$ Statistical Methodology, Novartis Pharma AG, Basel, Switzerland \\
    \small{$^{5}$ Section for Medical Statistics, Center for Medical Statistics, Informatics, and
Intelligent Systems,}  \\ \small Medical University of Vienna, Vienna, Austria 
}

	\maketitle
	
%


\parindent 0cm

\maketitle
\begin{abstract}
    A common problem in clinical trials is to test whether the effect of an explanatory variable on a response of interest 
    is similar between two groups, e.g. patient or treatment groups. In this regard, similarity is defined as equivalence up to a pre-specified threshold that denotes an acceptable deviation between the two groups. 
    This issue is typically tackled by assessing if the explanatory variable's effect on the response is similar. This assessment is based on, for example, confidence intervals of differences or a suitable distance between two parametric regression models.
    Typically, these approaches build on the assumption of a univariate continuous or binary outcome variable. 
    However, multivariate outcomes, especially beyond the case of bivariate binary response, remain underexplored.
    This paper introduces an approach based on a generalised joint regression framework exploiting the Gaussian copula.
    Compared to existing methods, our approach accommodates various outcome variable scales, such as continuous, binary, categorical, and ordinal, including mixed outcomes in multi-dimensional spaces.
    We demonstrate the validity of this approach through a simulation study and an efficacy-toxicity case study, hence highlighting its practical relevance.
    

\end{abstract}

\section{Introduction} \label{Intro}
A common challenge in applied research, especially in clinical trials, is determining whether an explanatory variable's effect on a response variable is equivalent or similar across different groups 
\citep[see, e.g.,][]{Otto,Jhee}. In this context, similarity is defined as equivalence up to a \emph{similarity threshold} value. 
Equivalence tests are widely used in various fields, particularly to determine if a treatment has comparable effects in different groups, based, for instance, on gender, age or treatments, just to mention a few.
Moreover, they are commonly used to investigate whether two formulations of a drug have nearly the same effect and are hence considered to be interchangeable, the key question of bioequivalence studies  \citep[e.g.,][]{bioequivalence}.



One usually assesses the question of similarity by testing whether the (marginal) effects of covariates on a response variable are similar among the groups, either based on confidence interval inclusion \citep{Liu2009, Gsteiger2011, Bretz2018} or using various distance measures as test statistics \citep{Moellenhoff2018, Dette2018}. 
These approaches assume a univariate continuous outcome variable which, as outlined by \citet{Moellenhoff2021}, might not be appropriate in many applications. 
On the one hand, the outcome might be, e.g., binary, categorical or ordinal. 
On the other hand, multivariate (often bivariate) outcomes arise, such as when analysing the efficacy and toxicity of a drug \citep[e.g.,][]{Jhee} which cannot be assumed to be independent of each other and, therefore, need to be modelled jointly. 

There are different approaches to jointly model multivariate outcome variables based on copulae \citep{Sklar}. 
\citet{Tao} proposed to use Archimedean copulae for such models, while \citet{Moellenhoff2021} suggested the Gumbel model \citep{Murtaugh, Heise} based on the Farlie-Gumbel-Morgenstern copula, which also belongs to the class of Archimedean copulae. In contrast, other authors employed elliptical copulae, especially the Gaussian which was adopted by \citet{deLeon2011} for regression models with bivariate mixed outcomes, and \citet{Weihsueh} for bivariate binary outcomes. The Gaussian copula is rather flexible for practical modelling: although it assumes linear dependence, it easily generalises to more than two dimensions and neatly characterises multivariate dependence through the covariance matrix \citep{Joe}. It also makes it possible to combine several types of variables (e.g., continuous, binary, categorical, continuous non-negative, ordinal) following various distributions. In different applied contexts, \citet{Radice2016} introduced bivariate models with binary margins, which were then generalised by \citet{Filippou} to the multivariate (specifically, trivariate) case. \citet{Marra2017} introduced bivariate copula models with continuous margins and \citet{Klein} additionally developed models for mixed responses (binary and continuous). The aforementioned models belong to the class of \emph{generalised joint regression models} implemented in the \texttt{R}-package \texttt{GJRM} \citep{GJRM}. Note that \texttt{GJRM} allows for many more modelling options than those mentioned here \citep[e.g.,][]{MRJASA,MRZ}.


To generalise a distance based similarity test for associated bivariate binary responses, \citet{Moellenhoff2021} adopted a copula approach to jointly model the efficacy and toxicity of a drug. 
However, the method proposed in this paper is more flexible, allowing for arbitrary dimensions and various types (including mixed) of outcome variables.
The proposal is based on the generalised joint regression framework with Gaussian copula and:
\begin{itemize}
     \item can be applied to multivariate responses of any size,
     \item accommodates various outcome types, including continuous, binary, and ordinal,
     \item adopts another type of test statistic that leads to a higher statistical power,
     \item addresses the problem of increasing type I error rates with increasing sample size, observed by \citet{Moellenhoff2021}.
\end{itemize}

The paper is structured as follows: 
In Section \ref{Modelling}, the modelling framework, based on generalised joint regression models with Gaussian copula, is succinctly discussed. 
In Section \ref{Test}, the new testing approach is introduced. Type I and II error rates for three relevant applied cases (bivariate binary, bivariate continuous and bivariate mixed outcomes) are studied in Section \ref{Sim}.
Section \ref{Empirical} illustrates the method using clinical trial data.

\section{Copula regression models} \label{Modelling}

\subsection{Regression structures}

Let $i=1,...,n$ be the observation index, where $n$ denotes the sample size. For a univariate outcome, the adopted modelling approach relies on the flexible regression structure 
$$\mu_i=m(x_i, \boldsymbol{\theta}),$$
where $\mu_i=\EE(y_i)$, $y_i \in \mathcal{Y} \subseteq \RR$ denotes the response variable within the set $\mathcal{Y}$ of all possible outcomes, $x_i \in \mathcal{X} \subseteq \RR$ is a deterministic explanatory variable, $m(\cdot)$ is a function modelling the effect of $x_i$ on $y_i$ via a regression curve, and $\boldsymbol{\theta} \in \RR^{\dim(\boldsymbol{\theta})}$ the related parameter vector. The function $m(\cdot)$ can be linear or non-linear, as illustrated in later sections.
We assume that $m(\cdot)$ is continuous, consequently resulting in continuous distances among model curves.


In the following, we are interested in comparing the effect of the variable $x_i$ on $y_i$ for two separate groups. 
This requires an additional group index $l=1,2$. 
Consequently, we observe outcomes $y_i^{(l)}$, $i=1,\ldots,n^{(l)}$, $l=1,2$, and regressions curves
$$\mu_i^{(1)}=m^{(1)}(x^{(1)}_i, \boldsymbol{\theta}^{(1)}) \quad \text{and} \quad \mu_i^{(2)}=m^{(2)}(x^{(2)}_i, \boldsymbol{\theta}^{(2)}).$$
For a multivariate outcome $\boldsymbol{y_i^{(l)}}=(y_{i1}^{(l)},\ldots,y_{iK}^{(l)})$, this generalises to 
$$
\boldsymbol{\mu}_i^{(l)}=\boldsymbol{m}^{(l)}(x^{(l)}_i, \boldsymbol{\theta}^{(l)}) \quad \Leftrightarrow \quad 
\begin{pmatrix}
    \mu_{i1}^{(l)} \\
    \vdots \\
    \mu_{iK}^{(l)}
\end{pmatrix} = 
\begin{pmatrix}
    m_1^{(l)}(x^{(l)}_i, \boldsymbol{\theta}_1^{(l)}) \\
    \vdots \\
    m_K^{(l)}(x^{(l)}_i, \boldsymbol{\theta}_K^{(l)})
\end{pmatrix},  \quad  l = 1, 2,
$$
with outcome dimension $K \in \NN^+$ and  group index $l$.
In dose-response studies with efficacy-toxicity outcomes, we have that $K=2$, the outcomes $y_{i1}^{(l)}$ and  $y_{i2}^{(l)}$ express the  efficacy and the toxicity, respectively, and the explanatory variable $x_i^{(l)}$ describes the dose for patient $i$ in group $l$, $i=1,\ldots, n^{(l)}$, $l=1,2$. We thus have regression structures $m_1^{(1)}(x_i^{(1)}, \boldsymbol{\theta}_1^{(1)})$ and $m_2^{(1)}(x_i^{(1)}, \boldsymbol{\theta}_2^{(1)})$ modelling the efficacy and toxicity for group 1, and $m_1^{(2)}(x_i^{(2)}, \boldsymbol{\theta}_1^{(2)})$ and $m_2^{(2)}(x_i^{(2)}, \boldsymbol{\theta}_2^{(2)})$ for group 2, respectively.

In general, since the $K$ responses are assumed to be dependent, the models have to be estimated jointly as described in the following section. 

\subsection{Copulae} \label{copula}

Copulae can be used 
to characterise the multivariate distribution of the response variables $y_{i1}^{(l)}, ..., y_{iK}^{(l)}, \, l = 1,2$. 
Specifically, for a $K$-dimensional distribution with cumulative distribution function (cdf) $\boldsymbol{F}$ and univariate marginal cdfs $F_1,...,F_K$ following uniforms on $[0,1]$, the copula $C:[0,1]^K \rightarrow [0,1]$ and $\boldsymbol{F}$ are linked as follows \citep{Sklar}
$$
\boldsymbol{F}(y_1,...,y_k) = C(F_{1}(y_1), ..., F_{K}(y_K)),
$$
where group index $l$ has been omitted for simplicity. We refer to \citet{Joe} and \citet{Trivedi}, for comprehensive introductions to copulae.

 Commonly used classes of copulae include the Archimedean copulae (which encompasses the Gumbel, Frank and Clayton) and the meta-elliptical copulae (which includes the Gaussian). The choice of copula often depends on the specific application and its modelling requirements. For the purpose of the present work, we adopt the Gaussian copula which, as mentioned in the Introduction, offers the required flexibility and generality in modelling the multivariate dependence structure of the variables of interest \citep{Joe}. The Gaussian copula can be generically defined as
\begin{align*}
    \boldsymbol{F}(y_{i1},...,y_{iK}) & = C(F_{1}(y_{i1}), ..., F_{K}(y_{iK})) = \Phi_K(\Phi^{-1}(F_{1}(y_{i1})), ..., \Phi^{-1}(F_{K}(y_{iK})), \boldsymbol{\Gamma}),
\end{align*}
or, in the bivariate case, as
$$
\boldsymbol{F}(y_{i1},y_{i2}) = \Phi_2(\Phi^{-1}(F_{1}(y_{i1})), \Phi^{-1}(F_{2}(y_{i2})),\rho),
$$
where $\Phi_K$ is the cdf of the $K$ dimensional multivariate Gaussian distribution, $\Phi^{-1}$ is the quantile function of the univariate Gaussian, $F_1(y_{i1}), ..., F_K(y_{iK})$ are the cdfs of the marginal distributions, $\boldsymbol{\Gamma} = \textbf{Cor}(y_{i1}, ..., y_{iK})$ and $\rho = \text{Cor}(y_{i1}, y_{i2})$. Note that in the above, we have suppressed $\boldsymbol{\theta}_k$ from the related marginal cdf for notational convenience. 


\subsection{Log-likelihoods} \label{estimation}

The structure of the log-likelihood function to employ in model fitting depends on the marginals considered in the analysis. 
The general likelihood theory of the K-dimensional case is given by \cite{song}. However, for simplicity 
of exposition, we report the functions considered in the applied cases of this paper: bivariate binary, bivariate continuous and bivariate mixed outcomes. For the same reason, we drop index $l$ and $\boldsymbol{\theta}_1$ and $\boldsymbol{\theta}_2$ from the marginal cdfs.



The log-likelihood for bivariate continuous outcomes is given by \citep{Marra2017} 
$$
\ell(\boldsymbol{\theta}_1, \boldsymbol{\theta}_2, \rho) = 
\sum_{i=1}^n  \left( \log(c(F_{1}(y_{i1}), F_{2}(y_{i2}))) + \log(f_1(y_{i1})) + \log(f_2(y_{i2}))
\right),
$$
where the copula density is defined as 
$$
c(F_{1}(y_{i1}), F_{2}(y_{i2})) = 
\frac{\partial^2 C(F_1(y_{i1}), F_2(y_{i2}))}{\partial F_1(y_{i1}) \partial F_2(y_{i2})}
$$
and $f_1$ and $f_2$ are marginal densities.
For bivariate mixed outcomes, with $y_{i1}$ binary and $y_{i2}$ continuous, the log-likelihood is \citep{Klein} 
$$\ell(\boldsymbol{\theta}_1, \boldsymbol{\theta}_2, \rho) = 
\sum_{i=1}^n  \left(
(1-y_{i1})\log(F_{1|2}(0|y_{i2})) + y_{i1}\log(1- F_{1|2}(0|y_{i2})) + \log(f_2(y_{i2}))
\right),$$
where
$$
F_{1|2}(0|y_{i2}) = \frac{\partial C(F_1(0), F_2(y_{i2}))}{\partial F_2(y_{i2})}
$$
and $F_1(0) = \Prob(y_{i1}=0)$. 

When both outcomes are binary, the log-likelihood is \citep{Radice2016} 
$$
\ell(\boldsymbol{\theta}_1, \boldsymbol{\theta}_2, \rho) = 
\sum_{i=1}^n  \left( 
y_{i1}  y_{i2} \log p_{11i} + 
y_{i1}  (1 - y_{i2}) \log p_{10i} +
(1 - y_{i1})  y_{i2} \log p_{01i} + 
(1 - y_{i1}) (1 - y_{i2}) \log p_{00i} 
\right),
$$
where
$$
p_{11i} = \Prob(y_{i1}=1, y_{i2}=1)
= C(\Prob(y_{i1}=1), \Prob(y_{i2}=1)),
$$
$p_{10i} = \Prob(y_{i1}=1) - \Prob(y_{i1}=1, y_{i2}=1)$, 
$p_{01i} = \Prob(y_{i2}=1) - \Prob(y_{i1}=1, y_{i2}=1)$ and 
$p_{00i} = 1- ( \Prob(y_{i1}=1) + \Prob(y_{i2}=1) - \Prob(y_{i1}=1, y_{i2}=1))$.

As explained earlier, for this work, we specify function $C$ using the Gaussian copula. Regarding the margins, we employ the Bernoulli distribution (with logit, probit or c-log-log link)
when the outcome is binary, whereas the normal, logistic, or another distribution 
can be utilised 
for a continuous response.

We achieve model fitting via the \texttt{R}-package \texttt{GJRM} \citep{GJRM} whose parameter estimation is based on an efficient and stable implementation of the trust region algorithm.

\section{Testing for similarity of multivariate non-independent responses} \label{Test}
\subsection{Hypotheses} \label{marginalHyp}
One approach to assess similarity of two curves, $m^{(1)}$ and $m^{(2)}$ in the univariate case is based on the maximum absolute deviation between them. 
In this case, the hypotheses are
\begin{equation}
\label{H0}
    H_0: \max_{x \in \mathcal{X}} |m^{(1)}(x, \boldsymbol{\theta}^{(1)}) - m^{(2)}(x, \boldsymbol{\theta}^{(2)})| \geq \epsilon \quad \text{ vs.} \quad H_1: \max_{x \in \mathcal{X}} |m^{(1)}(x, \boldsymbol{\theta}^{(1)}) - m^{(2)}(x, \boldsymbol{\theta}^{(2)})| < \epsilon,
\end{equation}
where $\epsilon$ is a prespecified threshold for similarity \citep{Moellenhoff2021, Dette2018}. 
Rejecting the null hypothesis suggests that, for a given significance level, the curves are similar since their distance is lower than the threshold value. 

For multivariate responses, there are several possibilities to formulate hypotheses for (joint) similarity. An intuitive approach, which generalises the approach of \citet{Moellenhoff2021} for the bivariate case, would be testing for (joint) similarity of all the curves associated with the $K$ outcomes. Formally, this leads to testing the hypotheses
\begin{equation}
\label{H1}
\begin{array} {ccc}
    H_0:  d_k \geq  \epsilon_{k} \text{ for at least one } k \in \{1,...,K\} \quad
    \text{vs.} \quad
    H_1: d_k < \epsilon_{k} \; \forall \; k \in \{1,...,K\},
\end{array} 
\end{equation}
%
where 
$$
d_k=\max_{x \in \mathcal{X}}|m_{k}^{(1)}(x, \boldsymbol{\theta}_k^{(1)}) - m_{k}^{(2)}(x, \boldsymbol{\theta}_k^{(2)})|, \, k=1,...,K,
$$
denotes the maximum absolute deviation between the curves describing the $k$th response and $\epsilon_k$ the corresponding similarity threshold. Since the alternative hypothesis stated in \eqref{H1} is expressed as an intersection of sub-hypotheses of similarity for each of the $K$ outcomes, \citet{Moellenhoff2021} suggested to test all of these $K$  sub-hypotheses
\begin{equation}
\label{H2}
\begin{array} {ccc}
    H_0^{(1)}: d_1 \geq \epsilon_1 \quad  \text{vs.} \quad H_1: d_1 < \epsilon_1\\
     \vdots \\
    H_0^{(K)}: d_K \geq \epsilon_K \quad \text{vs.} \quad H_1: d_K < \epsilon_K
\end{array} 
\end{equation}
individually. According to the intersection union principle \citep{Berger1982}, the global null hypothesis \eqref{H1} is then rejected if all of these individual hypotheses are rejected. This procedure guarantees an $\alpha$-level test if all individual tests are of size $\alpha$. However, such an approach is known to be quite conservative, especially for small sample sizes or a large number of individual tests (i.e., for a large $K$) \citep{Moellenhoff2021}.

To this end, we propose an alternative testing procedure, which we will call the \emph{maximum of maxima} approach. This is based on a different type of test statistic, that is $d_{max}:=\max\{d_1, ..., d_K\}$. The basic idea of this approach is that if the largest difference is sufficiently small then all the other differences are small enough too. 
This approach is similar to the one used by \citet{Moellenhoff2024} to jointly test for similarity of more than one transition intensity in a competing risks model.
The corresponding hypotheses are then given by
\begin{equation}
    H_0: d_{max} \geq \epsilon \quad \text{vs.} \quad H_1: d_{max} < \epsilon,
    \label{Hmax}
\end{equation} 
where $\epsilon$ now represents a global similarity threshold, denoted by $\epsilon=\epsilon_1=...=\epsilon_K$. In general, the individual thresholds $\epsilon_k, \, k = 1,..., K,$ may vary across the $K$ outcomes. 
However, by adding a data transformation step to the analysis, it is still possible to incorporate unequal individual thresholds in many cases. For continuous outcomes, one can achieve this for example by linear rescaling.

\subsection{Testing procedure}  \label{marginalAlg}

To test the hypotheses \eqref{Hmax}, we propose a parametric bootstrap approach similar to Algorithm 1 of \citet{Moellenhoff2021} and to the method of \citet{Dette2018}. Conducting parametric bootstrap requires the simulation of multivariate correlated outcomes. 
For multivariate binary outcomes, data generation is based on the algorithm of \citet{EP}. For the continuous case, one can just sample from a multivariate normal distribution. For multivariate mixed outcomes, we employ the algorithm of \citet{Demirtas}. 
%


\begin{algorithm}
	\label{alg1} { \normalfont 
 $ $
  \begin{itemize}
			\item[(1)] 
   Obtain, via MLE, $\hat{\boldsymbol{\theta}}_{k}^{(l)}, \, l = 1,2, \, k=1,...,K
            $, by maximising for each group the relevant log-likelihood (see Section \eqref{estimation}). The test statistic is calculated as            $$
            \hat{d}_{max} = \max \{\hat d_1, ...,\hat d_K \},
            $$ where $$
            \hat d_k=\max_{x \in \mathcal{X}}|m_{k}^{(1)}(x, \hat{\boldsymbol{\theta}}_{k}^{(1)}) - m_{k}^{(2)}(x, \hat{\boldsymbol{\theta}}_{k}^{(2)})|, \quad k=1,...,K.
            $$
			\item[(2)]  
			To approximate the null distribution, define estimators for parameter vectors $\boldsymbol{\theta}_{k}^{(l)}, \, l = 1,2, \, k=1,...,K$, so that the corresponding curves fulfil the null hypothesis in \eqref{Hmax}. That is,
			\begin{equation} \label{MLcons}
			\hat{\hat{\boldsymbol{\theta}}}_{k}^{(l)} = \left\{
			\begin{array} {ccc}
			\hat{\boldsymbol{\theta}}_{k}^{(l)} & \mbox{if} & \hat{d}_{max} \geq \epsilon \nonumber \\
		   \bar{\boldsymbol{\theta}}_{k}^{(l)} & \mbox{if} & \hat{d}_{max} < \epsilon
			\end{array}  \right. \quad l = 1,2, \, k=1,...,K,
			\end{equation}
			where $ \bar{\boldsymbol{\theta}}_{k}^{(l)}$ maximises the same objective function as $
            \hat{\boldsymbol{\theta}}_{k}^{(l)}, \, l = 1,2, \, k=1,...,K$ does, but 
			under the constraint
			\begin{equation}\label{constr}
			d_{max} = \epsilon.
			\end{equation}
Technically, we discretise the range, ${\mathcal X}$, of the explanatory variable to make the optimisation feasible. The constrained problem is solved using the augmented Lagrangian minimisation algorithm via function \texttt{auglag()} in the \texttt{R} package alabama \citep{alabama}.
				\item[(3)]
				Execute the following steps:
			\begin{itemize}
				\item[(i)] Obtain bootstrap samples under the null hypothesis in \eqref{Hmax} by generating data according to the model parameters $\hat{\hat{\boldsymbol{\theta}}}_{k}^{(l)}, \, l = 1,2, \, k=1,...,K$. This is achieved by obtaining parameter estimates for the marginal distributions and correlations and then feeding them into the data generation algorithms introduced above. 
				\item[(ii)] From the bootstrap samples, calculate the MLE $\hat{\boldsymbol{\theta}}_{k}^{(l)*}$ as in step $(1)$ and the test statistic
				\begin{equation}
				    \label{boot}
        \hat{d}^*_{max} = \max \{\hat{d}^*_1, ...,\hat{d}^*_K \},
        \end{equation}
        where $$
        \hat{d}_k^* = \max_{x \in \mathcal{X}}|m_{k}^{(1)}(x, \hat{\boldsymbol{\theta}}_{k}^{(1)*}) - m_{k}^{(2)}(x, \hat{\boldsymbol{\theta}}_{k}^{(2)*})|, \, k = 1,...,K.
            $$
		\item[(iii)] Repeat steps (i) and (ii) $n_{boot}$ times to generate replicates $\hat{d}^*_{max,1}, \dots, \hat{d}^*_{max,{(n_{boot})}}$ of $\hat{d}^*_{max}$.
            Let $\hat{d}^*_{max, (1)} \leq \ldots \leq \hat{d}^*_{max, (n_{boot})}$
			denote the corresponding order statistic. The estimator of the $\alpha$-quantile of the distribution of $\hat{d}^*_{max}$
			is given by $\hat{d}^*_{max,(\lfloor n_{boot} \alpha \rfloor )}$. 
            Reject the null hypothesis in \eqref{Hmax} 
				and assess similarity based on
			\begin{equation} \label{testInf}
			\hat{d}_{max} < \hat{d}^*_{max,(\lfloor n_{boot} \alpha \rfloor )}.
			\end{equation}
			Alternatively, obtain the $p$-value based on $\hat F_{n_{boot}}(\hat{d}_{max}) = {\frac{1}{n_{boot}}} \sum_{i=1}^{n_{boot}} \eins (\hat{d}^*_{max,i} \leq \hat{d}_{max})$ and reject the null hypothesis in \eqref{Hmax} if $\hat F_{n_{boot}}(\hat{d}_{max}) < \alpha$ for a pre-specified significance level $\alpha$, where $\hat F_{n_{boot}}$ denotes the empirical cumulative distribution function of the bootstrap sample.
	\end{itemize}			\end{itemize}
}
\end{algorithm}

The test presented in Algorithm \ref{alg1} has asymptotic level $\alpha$ and is consistent. That is, under the null hypothesis in \eqref{Hmax}, $
\limsup_{n \rightarrow\infty}\mathbb{P}\big(
  \hat{d}_{max}\leq \hat{d}^*_{max,(\lfloor n_{boot} \alpha \rfloor )}
\big)\leq\alpha$ and, under the alternative, 
$\lim_{n\rightarrow\infty} \mathbb{P}
\big(
   \hat{d}\leq \hat{d}^*_{max,(\lfloor n_{boot} \alpha \rfloor )}  \big) 
   =1$ for any $\alpha \in (0, 0.5)$. A formal proof of this can be directly obtained by transferring the proof given in \citet{Moellenhoff2024}, who investigate the same type of test statistic.  Precisely, it is based on
   the fact that the MLE $\hat{\boldsymbol{\theta}}_{k}^{(l)}$, $k=1,\ldots,K$, $l=1,2$, obtained by maximising the log-likelihoods given in Section \ref{estimation}, converge weakly to a normal distribution, 
   such that
   the proof of the general procedure of a constrained bootstrap given in \citet{Dette2018} can be adapted as described by \citet{Moellenhoff2024}.
   We investigate these properties for finite sample sizes in the following section.

\section{Finite sample properties}\label{Sim}

In this section, we investigate type I error rates and power of the proposed approach using Algorithm \ref{alg1}. Therefore, we simulate bivariate efficacy-toxicity outcomes as a function of dose, modelled by dose-response curves. 

For comparability with existing studies, particularly those in \citet{Moellenhoff2021}, the simulation setup for the bivariate binary case closely follows their scenarios. This includes the data generation routine and the levels of the explanatory variable $x^{(l)}_i$, set at specific dose levels $0, 0.1, 0.2, 0.5, 1, 1.5$ and $2$.
The simulation involves seven dose groups ($g=1,...,7$), with equal sample sizes $n_{g}^{(l)} \in \{7,14,21,28,50\}$, resulting in total sample sizes from $49$ to $350$ per group.
Note that we generate the data for each of the seven dose levels separately due to the algorithms' limitations in handling varying marginal probabilities/means, as detailed in Section \ref{marginalAlg}.

We keep the correlation parameter $\rho$ constant within each group $g$, leading to a different global correlation of the combined data \citep[see][for details of the relationship between group-wise and global correlations]{Dunlap}. Furthermore, we assume $\rho^{(1)}=\rho^{(2)}=\rho$, and employ the two different similarity threshold values, $\epsilon = 0.15$ and 0.2, introduced by \citet{Moellenhoff2021}. 
Given the computational costs related to the augmented Lagrangian minimisation algorithm and the data generation process, the study comprises 1000 simulation replicates and 300 bootstrap repetitions.

\subsection{Bivariate binary outcome}\label{binary-binary}
We adopt the same configurations as \citet{Moellenhoff2021}, employing Bernoulli marginals with logit links for both efficacy and toxicity, and 
$$
\boldsymbol{\theta}^{(l)}=
(\boldsymbol{\theta}_1^{(l)}, \boldsymbol{\theta}_2^{(l)}, \rho)=
(\beta_{01}^{(l)}, \beta_{11}^{(l)}, \beta_{02}^{(l)}, \beta_{12}^{(l)}, \rho), \quad l = 1,2,
$$ where $
\boldsymbol{\theta}^{(1)} = (-1, 2, -3, 3, \rho)
$.
To simulate type I error rates, we investigate $(d_1, d_2)\in\{(\epsilon, \epsilon),(0, \epsilon)\}$ for both $\epsilon=0.15$ and $\epsilon=0.2$, hence leading to 4 scenarios. Regarding the power, we investigate the three scenarios 
$(d_1, d_2) = (0.1, 0.1), (0.05, 0.05)$ and $(0, 0)$. The latter choice simulates the maximum power of the testing approach. The exact details of parameter combinations considered are shown in Table S1 of the \emph{supplementary material}. Finally, we investigate three different levels of group-wise correlations, $\rho= 0.1, 0.2$ and $0.3$.

Table \ref{bin_type1} shows the simulated type I error rates of the test implemented via Algorithm \ref{alg1}. We observe that, for $d_1=d_2 \approx \epsilon$, type I error rates are well below or very close the significance level of $\alpha = 0.05$. 
For the scenarios with $\min\{d_1, d_2\} = 0$, we observe slightly inflated type I error rates, up to a maximum of $0.106$ for the smallest group size of $n_{g}^{(l)} = 7$ and $\epsilon=0.2$. However, as the group size increases, the type I error rates decrease and approach the desired level of $0.05$. Of note, the value of $\rho$ does not seem to be that influential in this regard.


\begin{table}[htpb]
  \centering
  \caption{Simulated type I error rates of the test proposed in Algorithm \ref{alg1} for bivariate binary outcomes and two different similarity thresholds $\epsilon$. The nominal level is $\alpha=0.05$.}
    \begin{tabular}{ccccccc}
    \hline
    $\epsilon$ & $\boldsymbol{\theta}^{(2)}$ & $(d_1, d_2)$ & $n_{g}^{(l)}$     & $\rho = 0.1$  & $\rho = 0.2$ & $\rho = 0.3$ \\
    \hline
    0.2 & (-2.4, 3.4, -1.8, 2.51, $\rho$) & (0.2, 0.2) & 7     & 0.031 & 0.037 & 0.038 \\
    0.2 & (-2.4, 3.4, -1.8, 2.51, $\rho$) & (0.2, 0.2) & 14    & 0.012 & 0.012 & 0.018 \\
    0.2 & (-2.4, 3.4, -1.8, 2.51, $\rho$) & (0.2, 0.2) & 21    & 0.013 & 0.006 & 0.012 \\
    0.2 & (-2.4, 3.4, -1.8, 2.51, $\rho$) & (0.2, 0.2) & 28    & 0.007 & 0.006 & 0.005 \\
    0.2 & (-2.4, 3.4, -1.8, 2.51, $\rho$) & (0.2, 0.2) & 50    & 0.006 & 0.009 & 0.004 \\
    0.2 & (-1, 2, -1.8, 2.51, $\rho$) & (0, 0.2) & 7    & 0.072 & 0.106 & 0.106 \\
    0.2 & (-1, 2, -1.8, 2.51, $\rho$) & (0, 0.2) & 14    & 0.084 & 0.100 & 0.089 \\
    0.2 & (-1, 2, -1.8, 2.51, $\rho$) & (0, 0.2) & 21    & 0.082 & 0.088 & 0.078 \\
    0.2 & (-1, 2, -1.8, 2.51, $\rho$) & (0, 0.2) & 28    & 0.064 & 0.070 & 0.078 \\
    0.2 & (-1, 2, -1.8, 2.51, $\rho$) & (0, 0.2) & 50    & 0.054 & 0.066 & 0.058 \\
    0.15  & (-2, 3.4,-2, 2.51, $\rho$)& (0.15, 0.15) & 7     & 0.057 & 0.051 & 0.058 \\
    0.15  & (-2, 3.4,-2, 2.51, $\rho$)& (0.15, 0.15) & 14    & 0.032 & 0.022 & 0.026 \\
    0.15  & (-2, 3.4,-2, 2.51, $\rho$)& (0.15, 0.15) & 21    & 0.021 & 0.022 & 0.020 \\
    0.15  & (-2, 3.4,-2, 2.51, $\rho$)& (0.15, 0.15) & 28    & 0.012 & 0.013 & 0.007 \\
    0.15  & (-2, 3.4,-2, 2.51, $\rho$)& (0.15, 0.15) & 50    & 0.013 & 0.010  & 0.008 \\
    0.15  & (-1, 2,-2, 2.51, $\rho$)& (0, 0.15) & 7     & 0.089 & 0.097 & 0.088 \\
    0.15  & (-1, 2,-2, 2.51, $\rho$)& (0, 0.15) & 14    & 0.085 & 0.077 & 0.087 \\
    0.15  & (-1, 2,-2, 2.51, $\rho$)& (0, 0.15) & 21    & 0.075 & 0.081 & 0.082 \\
    0.15  & (-1, 2,-2, 2.51, $\rho$)& (0, 0.15) & 28    & 0.062 & 0.068 & 0.088 \\
    0.15  & (-1, 2,-2, 2.51, $\rho$)& (0, 0.15) & 50    & 0.067 & 0.083 & 0.073 \\

\hline
    \end{tabular}%
  \label{bin_type1}%
\end{table}%

In comparison to \citet{Moellenhoff2021}, where type I error rates were predominantly close to zero, the results for our approach align more closely with the nominal level. This is in line with the theoretical arguments of Section \ref{marginalHyp}: the proposal is less conservative compared to testing based on the intersection union principle. However, for some configurations with high correlation, \cite{Moellenhoff2021} observed an inflation of the type I error rates as the sample size increases up to a value of $12.7 \%$. In contrast, the type I error rates decrease for increasing sample sizes when using our approach. In addition, 
the maximum type I error rate is $10.6 \%$ for our approach, which is considerably smaller than the $12.7 \%$ observed in \citet{Moellenhoff2021},
although we use a less conservative approach. This might suggests that the model based on the Gaussian copula outperforms the Gumbel model in this setting. 

\begin{figure}[t]
    \centering
    \begin{overpic}[scale=0.66]{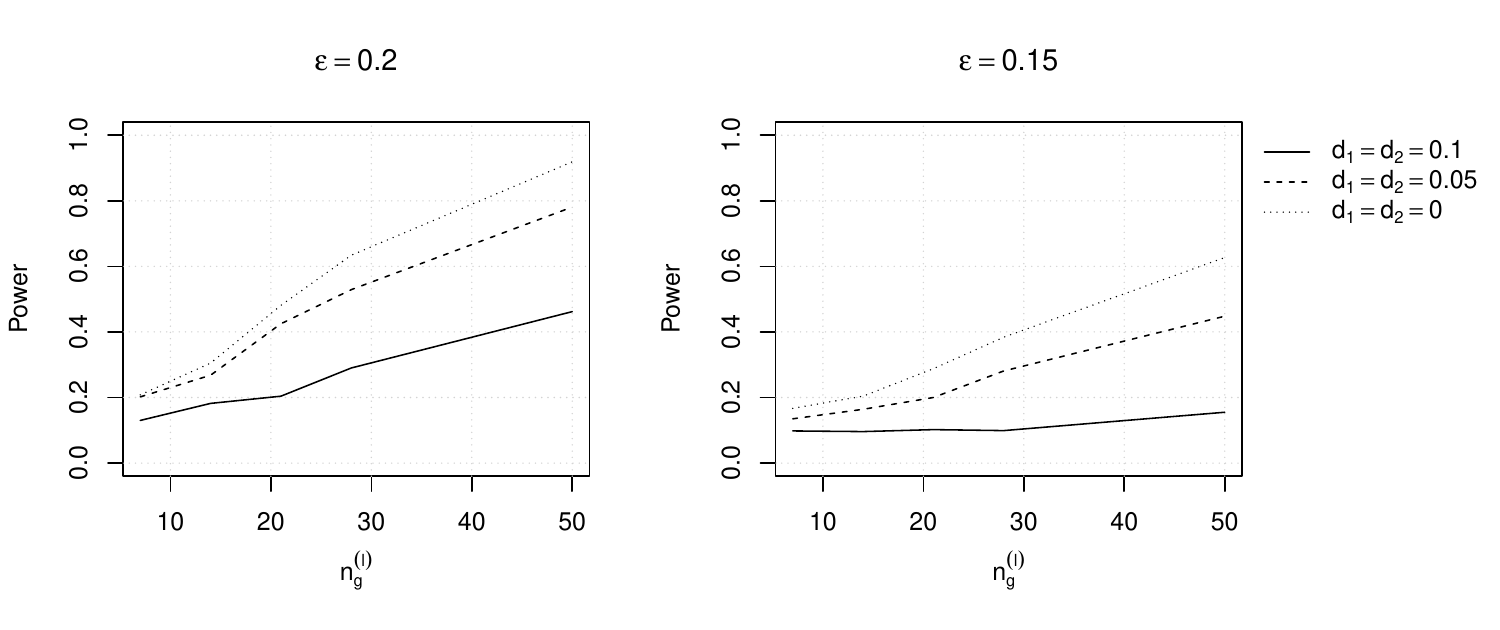}
        \put(0,39){(a)}
        \put(43,39){(b)}
    \end{overpic}
    \caption{Simulated power of the test proposed in Algorithm \ref{alg1} for bivariate binary outcomes for different sample sizes and $\rho = 0.2$. The three different scenarios are shown in terms of different line types. The nominal level is $\alpha=0.05$.}
    \label{binary_curves}
\end{figure}

Figure \ref{binary_curves} displays the simulated power as function of sample size for the different scenarios. 
As for the type I errors, the level of correlation has little effect on power, shown in detail for $\rho = 0.2$ in Figure 1 (complete results in Table S2 of the supplementary material).
Our testing approach shows increasing power with larger sample sizes, converging to 1 in all scenarios.
The highest power of $0.919$ is observed for $(d_1,d_2)=(0,0)$ with $\epsilon = 0.2$, $\rho = 0.2$ and $n_{g}^{(l)}=50$ (see Figure \ref{binary_curves}(a)). 
For a medium sample size ($n_{g}^{(l)} \in \{21, 28\}$), the power is between $0.214$ and $0.650$ for $\epsilon = 0.2$, and between $0.095$ and $0.384$ for $\epsilon = 0.15$, respectively. Finally, when considering small sample sizes ($n_{g}^{(l)} \in \{7, 14\}$), our model still achieves reasonable power, with values from $0.128$ to $0.334$ for $\epsilon = 0.2$, and from $0.086$ to $0.204$ for $\epsilon = 0.15$.
Compared to \citet{Moellenhoff2021}, our method demonstrates similar high power for large samples but considerably higher power for small and medium samples (exceeding in some cases by over fivefold). This, again, highlights that the proposed approach is less conservative relative to approaches based on the intersection union principle.


\subsection{Bivariate continuous outcome} \label{cont-cont}

\begin{figure}[t]
    \centering
    \includegraphics[scale = 0.7]{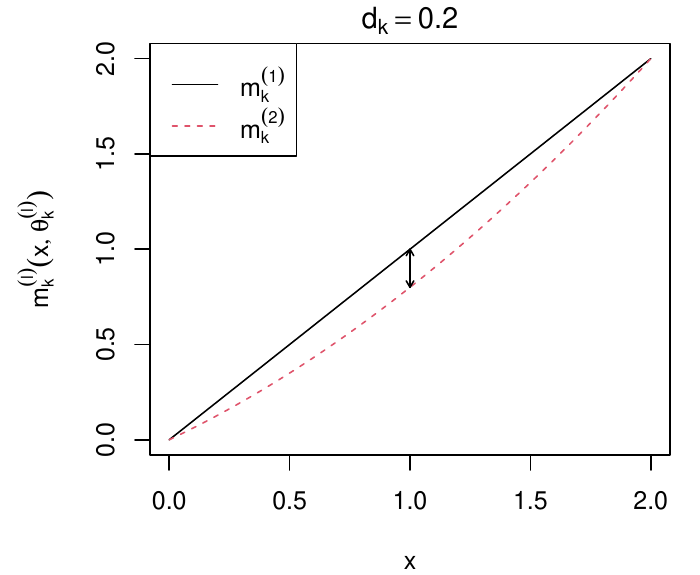}
    \caption{Visualisation of the curves \eqref{scen4.2} for the two groups $l=1,2$, where the maximum distance $d_k=0.2$ is observed for $x=1$ and corresponds to the length of the arrow.}
    \label{contin-contin_curves}
\end{figure}

In the case of bivariate continuous outcomes, we adopt the set up of \citet{Bretz2018}, i.e. a linear dose-response model $$m_k^{(1)}(x^{(1)}_i, \boldsymbol{\theta}_k^{(1)})= 
\beta_{0k}^{(1)} + \beta_{1k}^{(1)} x_i^{(1)}$$ for the first group, and a quadratic model
$$m_k^{(2)}(x^{(2)}_i, \boldsymbol{\theta}_k^{(2)})= 
\beta_{0k}^{(2)} + \beta_{1k}^{(2)} x^{(2)}_i + \beta_{2k}^{(2)} \left( x^{(2)}_i \right) ^2 $$
for the second group, $i=1,\ldots n_l$, $l=1,2$, $k=1,2$. The related full parameter vectors are given by
$$
\boldsymbol{\theta}^{(1)}=
(\boldsymbol{\theta}_1^{(1)}, \boldsymbol{\theta}_2^{(1)}, \sigma, \rho)=
(\beta_{01}^{(1)}, \beta_{11}^{(1)}, \beta_{02}^{(1)}, \beta_{12}^{(1)}, \sigma, \rho)
$$
and
$$
\boldsymbol{\theta}^{(2)}=
(\boldsymbol{\theta}_1^{(2)}, \boldsymbol{\theta}_2^{(2)}, \sigma, \rho)=
(\beta_{01}^{(2)}, \beta_{11}^{(2)}, \beta_{21}^{(2)},\beta_{02}^{(2)}, \beta_{12}^{(2)}, \beta_{22}^{(2)}, \sigma, \rho).
$$
For consistency with Section \ref{binary-binary}, we transform the model such that it applies to the dose range of $x_i \in [0,2]$; this is achieved by setting $
\beta_{0k}^{(1)} = 0, 
\beta_{1k}^{(1)} = 1,
\beta_{0k}^{(2)} = 0,
\beta_{1k}^{(2)} = (1 - 2d_k) $ and $ 
\beta_{2k}^{(2)} = d_k
$, so that
\be\label{scen4.2}
m_k^{(1)}(x^{(1)}_i, \boldsymbol{\theta}_k^{(1)})= x^{(1)}_i \quad \text{and} \quad
m_k^{(2)}(x^{(2)}_i, \boldsymbol{\theta}_k^{(2)})= (1 - 2d_k) x^{(2)} + d_k \left( x^{(2)}_i \right)^2,
\quad k=1,2,\ee
where $d_k$ is the corresponding distance between the curves. 
This leads to $
\boldsymbol{\theta}^{(1)} = (0,1,0,1, \sigma, \rho)
$ for all scenarios. 
The curves coincide at the boundary doses $x^{(l)}_i=0$ and $x^{(l)}_i=2$, and the maximum difference $d_k$ occurs at a middle dose of $x^{(l)}_i=1$ (see Figure \eqref{contin-contin_curves} for an example when $d_k=0.2$).

As in Section \ref{binary-binary}, we assume equal correlation for both groups $(\rho= \rho^{(1)}=\rho^{(2)})$ and equal variances for the continuous variables across responses and groups, i.e. $\sigma= \sigma_1^{(1)}= \sigma_2^{(1)}= \sigma_1^{(2)}= \sigma_2^{(2)}$. The variance levels are chosen to be $\sigma^2 = 0.05, 0.1, 0.2$  such that the ratios $\epsilon/\sigma$ are similar to the ones chosen by \citet{Bretz2018}. We investigate the same seven scenarios as in Section \ref{binary-binary}, i.e. $(d_1, d_2)=(\epsilon, \epsilon)$ and $(0, \epsilon)$ with $\epsilon=0.15,0.2$ for the type I error rate simulation, and 
$(d_1, d_2) = (0.1, 0.1), (0.05, 0.05)$ and $(0, 0)$ for the power simulations. The complete parameter combinations are in Table S3 of the \emph{supplementary materials}.  


Table \ref{tab:cont_type1} shows the simulated type I error rates. As already observed for the bivariate binary case, the level of correlation has little impact on the results, hence we only report the type I error rates for the medium correlation level of $\rho = 0.2$ and refer the reader to Table S4 and S5 of the \emph{supplementary materials} for the complete set of results. 
For $d_1=d_2 \approx \epsilon$, Type I error rates closely align with the $5\%$ level across all configurations. Similar to the findings in Section 4.1, we note a slight inflation in Type I errors for $(d_1, d_2)=(0, \epsilon)$, consistent in magnitude with the binary outcomes. Notably, with lower variance, Type I error rates tend to decrease and align with the desired $5\%$ level as sample sizes increase. We partially observe a similar trend for higher variance levels.

\begin{table}[htb]
  \centering
  \caption{Simulated type I error rates of the test proposed in Algorithm \ref{alg1} for bivariate continuous outcomes specified in \eqref{scen4.2} with $\rho = 0.2$ and two different similarity thresholds $\epsilon$. The nominal level is $\alpha=0.05$.}
    \begin{tabular}{ccccccc}
    \hline
    $\epsilon$ & $\boldsymbol{\theta}^{(2)}$ & $(d_1, d_2)$ & $n_{g}^{(l)}$     & $\sigma^2 = 0.05$  & $\sigma^2 = 0.1$ & $\sigma^2 = 0.2$ \\
     \hline
    0.2   & $(0, 0.6,  0.2, 0, 0.6,  0.2, \sigma, \rho)$ & $(0.2, 0.2)$ & 7     & 0.044 & 0.037 & 0.045 \\
    0.2   & $(0, 0.6,  0.2, 0, 0.6,  0.2, \sigma, \rho)$ & $(0.2, 0.2)$ & 14    & 0.029 & 0.037 & 0.038 \\
    0.2   & $(0, 0.6,  0.2, 0, 0.6,  0.2, \sigma, \rho)$ & $(0.2, 0.2)$ & 21    & 0.016 & 0.021 & 0.039 \\
    0.2   & $(0, 0.6,  0.2, 0, 0.6,  0.2, \sigma, \rho)$ & $(0.2, 0.2)$ & 28    & 0.005 & 0.012 & 0.022 \\
    0.2   & $(0, 0.6,  0.2, 0, 0.6,  0.2, \sigma, \rho)$ & $(0.2, 0.2)$ & 50    & 0.013 & 0.012 & 0.018 \\
    0.2   & $(0, 1, 0, 0, 0.6, 0.2, \sigma, \rho)$ & $(0, 0.2)$ & 7     & 0.090 & 0.104 & 0.095 \\
    0.2   & $(0, 1, 0, 0, 0.6, 0.2, \sigma, \rho)$ & $(0, 0.2)$ & 14    & 0.064 & 0.087 & 0.089 \\
    0.2   & $(0, 1, 0, 0, 0.6, 0.2, \sigma, \rho)$ & $(0, 0.2)$ & 21    & 0.075 & 0.073 & 0.080 \\
    0.2   & $(0, 1, 0, 0, 0.6, 0.2, \sigma, \rho)$ & $(0, 0.2)$ & 28    & 0.054 & 0.072 & 0.086 \\
    0.2   & $(0, 1, 0, 0, 0.6, 0.2, \sigma, \rho)$ & $(0, 0.2)$ & 50    & 0.056 & 0.077 & 0.079 \\
    0.15  & $(0, 0.7, 0.15, 0, 0.7, 0.15, \sigma, \rho)$ & $(0.15, 0.15)$ & 7     & 0.044 & 0.055 & 0.07 \\
    0.15  & $(0, 0.7, 0.15, 0, 0.7, 0.15, \sigma, \rho)$ & $(0.15, 0.15)$ & 14    & 0.035 & 0.045 & 0.06 \\
    0.15  & $(0, 0.7, 0.15, 0, 0.7, 0.15, \sigma, \rho)$ & $(0.15, 0.15)$ & 21    & 0.018 & 0.044 & 0.043 \\
    0.15  & $(0, 0.7, 0.15, 0, 0.7, 0.15, \sigma, \rho)$ & $(0.15, 0.15)$ & 28    & 0.017 & 0.033 & 0.040 \\
    0.15  & $(0, 0.7, 0.15, 0, 0.7, 0.15, \sigma, \rho)$ & $(0.15, 0.15)$ & 50    & 0.014 & 0.021 & 0.034 \\
    0.15  & $(0, 1, 0, 0, 0.7, 0.15, \sigma, \rho)$ & $(0, 0.15)$ & 7     & 0.096 & 0.079 & 0.106 \\
    0.15  & $(0, 1, 0, 0, 0.7, 0.15, \sigma, \rho)$ & $(0, 0.15)$ & 14    & 0.092 & 0.099 & 0.096 \\
    0.15  & $(0, 1, 0, 0, 0.7, 0.15, \sigma, \rho)$ & $(0, 0.15)$ & 21    & 0.081 & 0.084 & 0.089 \\
    0.15  & $(0, 1, 0, 0, 0.7, 0.15, \sigma, \rho)$ & $(0, 0.15)$ & 28    & 0.065 & 0.083 & 0.079 \\
    0.15  & $(0, 1, 0, 0, 0.7, 0.15, \sigma, \rho)$ & $(0, 0.15)$ & 50    & 0.054 & 0.089 & 0.096 \\
     \hline
    \end{tabular}%
  \label{tab:cont_type1}%
\end{table}%

\begin{figure}[t]
    \centering
    \begin{overpic}[scale=0.66]{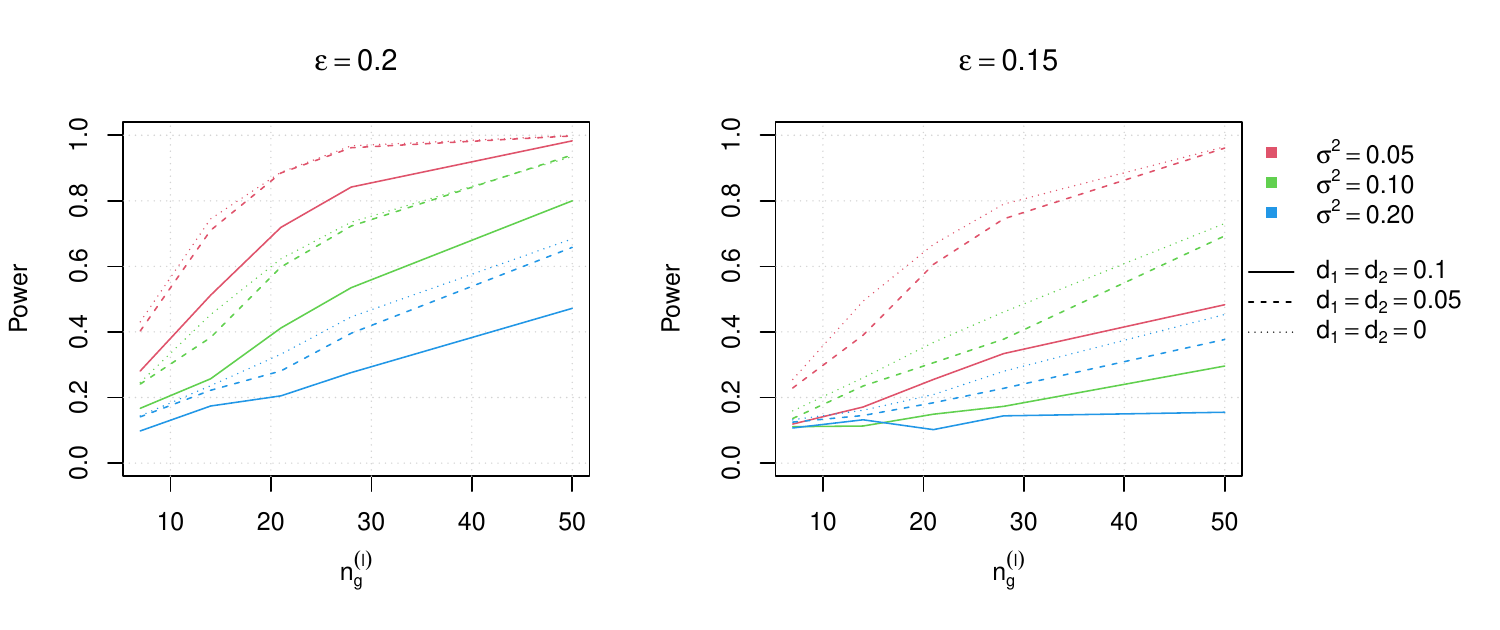}
        \put(0,39){(a)}
        \put(43,39){(b)}
    \end{overpic}
    \caption{Simulated power of the test proposed in Algorithm \ref{alg1} for  bivariate continuous outcomes for different sample sizes and $\rho = 0.2$. The different variance levels are shown in terms of colours and the three different scenarios are shown in terms of different line types. The nominal level is $\alpha=0.05$.}
    \label{contin-contin_power}
\end{figure}

Figure \ref{contin-contin_power} displays the simulated power. Again, the level of correlation has little influence, hence we focus on the medium correlation level of $\rho = 0.2$ (full results are in Table S6 - S8 of the \emph{supplementary materials}). The proposal achieves reasonable power.
For instance, for a medium sample size ($n_{g}^{(l)} \in \{21, 28\}$), we find a power between $0.205$ and $0.968$ for $\epsilon = 0.2$, and between $0.102$ and $0.790$ for $\epsilon = 0.15$. At small sample sizes, $n_{g}^{(l)} \in \{7, 14\}$, the approach still achieves satisfying power, reaching values from $0.098$ to $0.710$ for $\epsilon = 0.2$, and from $0.107$ to $0.494$ for $\epsilon = 0.15$. Finally, the power converges to $1$ for decreasing variance and increasing sample size.

\subsection{Bivariate mixed outcome}

For the case of bivariate mixed outcomes, we focus on binary and continuous responses and combine the scenarios considered in Sections \ref{binary-binary} and \ref{cont-cont}, corresponding to a bivariate efficacy-toxicity outcome. 

As in Section \ref{cont-cont}, we assume $\rho = \rho^{(1)} = \rho^{(2)}$, $\sigma = \sigma^{(1)} = \sigma^{(2)}$ and the same variance levels ($\sigma^2 = 0.05, 0.1$ and $0.2$). 
Different to the bivariate binary and continuous cases, $(d_1, d_2) = (0,\epsilon)$ and $(d_1, d_2) = (\epsilon,0)$ are not equivalent for bivariate mixed outcomes, hence we have to investigate them separately.
As a consequence, we observe nine different configurations of
$$
\boldsymbol{\theta}^{(1)}=
(\boldsymbol{\theta}_1^{(1)}, \boldsymbol{\theta}_2^{(1)}, \sigma, \rho)=
(\beta_{01}^{(1)}, \beta_{11}^{(1)}, \beta_{02}^{(1)}, \beta_{12}^{(1)}, \sigma, \rho)
$$
and
$$
\boldsymbol{\theta}^{(2)}=
(\boldsymbol{\theta}_1^{(2)}, \boldsymbol{\theta}_2^{(2)}, \sigma, \rho)=
(\beta_{01}^{(2)}, \beta_{11}^{(2)}, \beta_{21}^{(2)},\beta_{02}^{(2)}, \beta_{12}^{(2)}, \sigma, \rho),
$$
investigating $(d_1, d_2)=(\epsilon, \epsilon), (0, \epsilon)$ and $(\epsilon, 0)$ for $\epsilon=0.15,0.2$
for the type I error simulations and $(d_1, d_2) = (0.1, 0.1), (0.05, 0.05)$ and $(0, 0)$ for the power simulations, with $\boldsymbol{\theta}^{(1)}$ constantly held as $(0, 1, -1, 2, \sigma, \rho)$. 
The complete parameter combinations are in Table S9 of the \emph{supplementary materials}.


Table \ref{tab:mixed_type1} shows the simulated type I error rates for $\rho = 0.2$ 
and mirror findings from bivariate continuous outcomes, with $d_1=d_2 \approx \epsilon$ showing alignment with the 5 \% error level. Slight inflation is observed for $(d_1, d_2) = (0, \epsilon)$ and $(\epsilon,0)$. Variance, affecting only one outcome, seems to have a reduced impact on the type I error rates. Interestingly, $d_1 = 0$ tends to produce slightly higher type I error rates, aligning with observations from Section \ref{cont-cont}.
Results for $\rho = 0.1$ and $0.2$ are in Table S10 and S11 of the \emph{supplementary materials}.

\begin{table}[htbp]
  \centering
  \caption{Simulated type I error rates of the test proposed in Algorithm \ref{alg1} for bivariate mixed outcomes with $\rho = 0.2$ and two different similarity thresholds $\epsilon$. The nominal level is $\alpha=0.05$.}

    \begin{tabular}{ccccccc}
    \hline
    $\epsilon$ & $\boldsymbol{\theta}^{(2)}$ & $(d_1, d_2)$ & $n_{g}^{(l)}$     & $\sigma^2 = 0.05$  & $\sigma^2 = 0.1$ & $\sigma^2 = 0.2$ \\
    \hline
    0.2   & $(0, 0.6,  0.2, -2.4, 3.4 \sigma, \rho)$ & (0.2, 0.2) & 7     & 0.032 & 0.035 & 0.038 \\
    0.2   & $(0, 0.6,  0.2, -2.4, 3.4 \sigma, \rho)$ & (0.2, 0.2) & 14    & 0.022 & 0.031 & 0.03 \\
    0.2   & $(0, 0.6,  0.2, -2.4, 3.4 \sigma, \rho)$ & (0.2, 0.2) & 21    & 0.018 & 0.009 & 0.017 \\
    0.2   & $(0, 0.6,  0.2, -2.4, 3.4 \sigma, \rho)$ & (0.2, 0.2) & 28    & 0.014 & 0.016 & 0.019 \\
    0.2   & $(0, 0.6,  0.2, -2.4, 3.4 \sigma, \rho)$ & (0.2, 0.2) & 50    & 0.013 & 0.014 & 0.016 \\
    0.2   & $(0, 0.6,  0.2, -2, 2, \sigma, \rho)$ & (0.2, 0) & 7     & 0.067 & 0.076 & 0.083 \\
    0.2   & $(0, 0.6,  0.2, -2, 2, \sigma, \rho)$ & (0.2, 0) & 14    & 0.077 & 0.078 & 0.079 \\
    0.2   & $(0, 0.6,  0.2, -2, 2, \sigma, \rho)$ & (0.2, 0) & 21    & 0.070 & 0.071 & 0.103 \\
    0.2   & $(0, 0.6,  0.2, -2, 2, \sigma, \rho)$ & (0.2, 0) & 28    & 0.073 & 0.072 & 0.084 \\
    0.2   & $(0, 0.6,  0.2, -2, 2, \sigma, \rho)$ & (0.2, 0) & 50    & 0.049 & 0.057 & 0.069 \\
    0.2   & $(0, 1, 0, -2.4, 3.4, \sigma, \rho)$ & (0, 0.2) & 7     & 0.114 & 0.117 & 0.076 \\
    0.2   & $(0, 1, 0, -2.4, 3.4, \sigma, \rho)$ & (0, 0.2) & 14    & 0.082 & 0.075 & 0.070 \\
    0.2   & $(0, 1, 0, -2.4, 3.4, \sigma, \rho)$ & (0, 0.2) & 21    & 0.061 & 0.072 & 0.068 \\
    0.2   & $(0, 1, 0, -2.4, 3.4, \sigma, \rho)$ & (0, 0.2) & 28    & 0.059 & 0.055 & 0.062 \\
    0.2   & $(0, 1, 0, -2.4, 3.4, \sigma, \rho)$ & (0, 0.2) & 50    & 0.059 & 0.045 & 0.055 \\
    0.15  & $(0.7, 0.15, -2, 3.4, 0, \sigma, \rho)$ & (0.15, 0.15) & 7     & 0.037 & 0.036 & 0.051 \\
    0.15  & $(0.7, 0.15, -2, 3.4, 0, \sigma, \rho)$ & (0.15, 0.15) & 14    & 0.029 & 0.022 & 0.038 \\
    0.15  & $(0.7, 0.15, -2, 3.4, 0, \sigma, \rho)$ & (0.15, 0.15) & 21    & 0.023 & 0.024 & 0.034 \\
    0.15  & $(0.7, 0.15, -2, 3.4, 0, \sigma, \rho)$ & (0.15, 0.15) & 28    & 0.013 & 0.02  & 0.031 \\
    0.15  & $(0.7, 0.15, -2, 3.4, 0, \sigma, \rho)$ & (0.15, 0.15) & 50    & 0.018 & 0.009 & 0.019 \\
    0.15  & $(0.7, 0.15, -2, 2, 0, \sigma, \rho)$ & (0.15, 0) & 7     & 0.073 & 0.077 & 0.090 \\
    0.15  & $(0.7, 0.15, -2, 2, 0, \sigma, \rho)$ & (0.15, 0) & 14    & 0.075 & 0.077 & 0.082 \\
    0.15  & $(0.7, 0.15, -2, 2, 0, \sigma, \rho)$ & (0.15, 0) & 21    & 0.075 & 0.084 & 0.089 \\
    0.15  & $(0.7, 0.15, -2, 2, 0, \sigma, \rho)$ & (0.15, 0) & 28    & 0.048 & 0.069 & 0.084 \\
    0.15  & $(0.7, 0.15, -2, 2, 0, \sigma, \rho)$ & (0.15, 0) & 50    & 0.070 & 0.071 & 0.080 \\
    0.15  & $(1, 0, -2, 3.4, 0, \sigma, \rho)$ & (0, 0.15) & 7     & 0.097 & 0.091 & 0.092 \\
    0.15  & $(1, 0, -2, 3.4, 0, \sigma, \rho)$ & (0, 0.15) & 14    & 0.092 & 0.086 & 0.075 \\
    0.15  & $(1, 0, -2, 3.4, 0, \sigma, \rho)$ & (0, 0.15) & 21    & 0.076 & 0.099 & 0.076 \\
    0.15  & $(1, 0, -2, 3.4, 0, \sigma, \rho)$ & (0, 0.15) & 28    & 0.056 & 0.068 & 0.067 \\
    0.15  & $(1, 0, -2, 3.4, 0, \sigma, \rho)$ & (0, 0.15) & 50    & 0.045 & 0.051 & 0.073 \\
    \hline
    \end{tabular}%
  \label{tab:mixed_type1}%
\end{table}%

\begin{figure}[htb]
    \centering
    \begin{overpic}[scale=0.66]{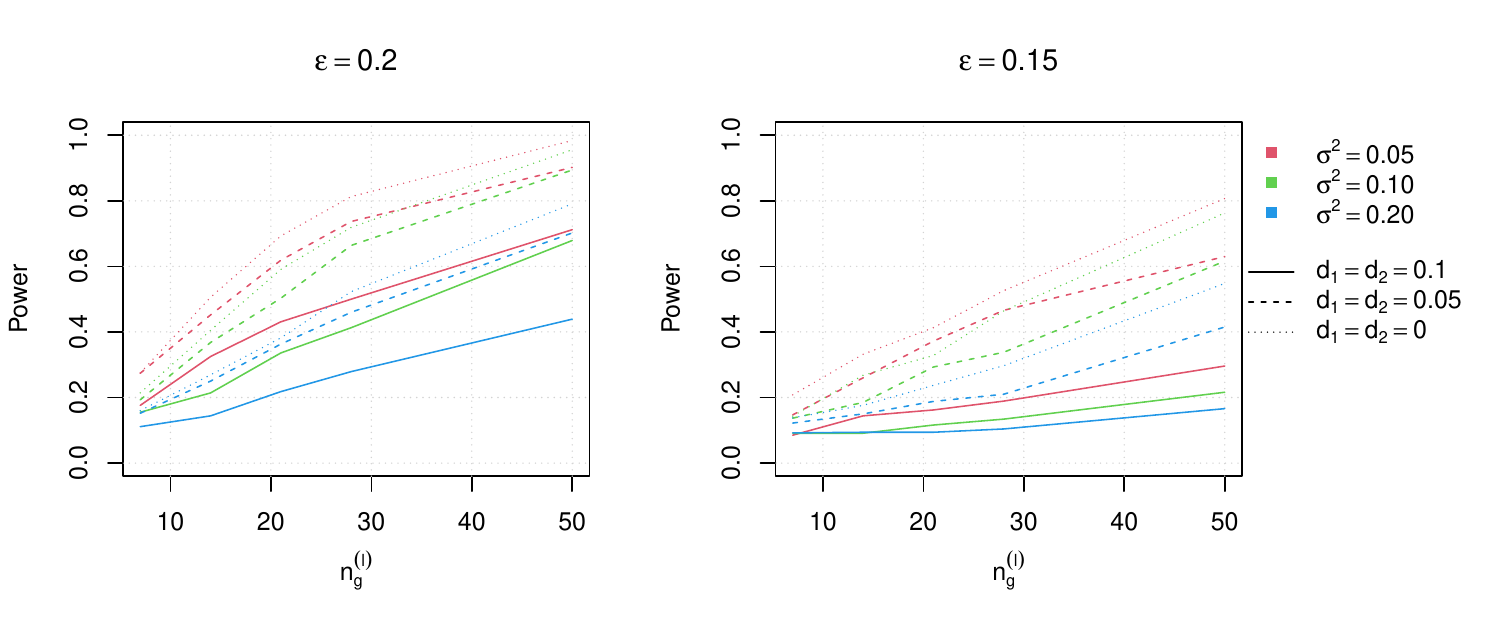}
        \put(0,39){(a)}
        \put(43,39){(b)}
    \end{overpic}
    \caption{Simulated power of the test proposed in Algorithm \ref{alg1} for bivariate mixed outcomes for different sample sizes and $\rho = 0.2$. The different variance levels are shown in terms of colours and the three different scenarios are shown in terms of different line types. The nominal level is $\alpha=0.05$.}
    \label{mixed_power}
\end{figure}

Figure \ref{mixed_power} shows the simulated power values for $\rho = 0.2$ 
with power generally increasing for smaller variances and larger sample sizes. We observe the maximum power of 0.984 for $(d_1, d_2) = (0,0)$, $\epsilon = 0.2$, and $n_{g}^{(l)}=50$. Medium sample sizes lead to power values ranging from $0.218$ to $0.813$ for $\epsilon = 0.2$, decreasing slightly for $\epsilon = 0.15$. Even with smaller sample sizes, our approach maintains reasonable power, reinforcing the robustness of our approach.

\section{Case study} \label{Empirical}

We illustrate the proposed methodology through a case study, inspired by \citet{Moellenhoff2021}.
The goal of this study was to investigate dental pain reduction of a nonsteroidal anti-inflammatory drug after the removal of two or more impacted third molar teeth. Specifically, interest was in assessing similarity with an already available marketed product with regard to a bivariate efficacy-toxicity outcome. 
For the purpose of the following analysis, we used a hypothetical data set, simulated according to real data, due to confidentiality reasons. A total of $n=300$ participants are evenly allocated to these 10 groups, resulting in $n_g^{(l)}=30$ patients per group.

Pain intensity is measured on an ordinal scale at baseline, and several times after the administration of a single dose. Even though the original scale is ordinal, the average over the repeated measurements can be modelled as a continuous variable. Besides the placebo, there are 4 dose levels for each drug ($g=1,...,5$), where the levels are 0.05, 0.20, 0.50, and 1 for the investigational drug and 0.10, 0.30, 0.60, and 1 for the marketed product, respectively. 
The actual doses are scaled to lie within the [0, 1] interval to maintain confidentiality. 
In order to incorporate different similarity thresholds $\epsilon_1 \ne \epsilon_2$ into the analysis, we linearly rescale the average pain reduction as suggested in Section \ref{marginalHyp}.

A binary toxicity variable indicates whether or not side effects (e.g., nausea and sedation after dosing) occur. Thus, the outcome of interest is a bivariate mixed outcome (continous-binary). The approach of \citet{Moellenhoff2021} is limited to binary variables, hence  the authors created a binary success variable for efficacy by comparing the average pain reduction to a clinical relevance threshold. Our approach is not restricted to binary efficacy outcomes so that we can consider the original average pain reduction on the continuous scale, allowing us to better exploit the available information in the analysis. Specifically, one can construct the dataset considered in \citet{Moellenhoff2021} from our dataset by using $0.5$ as the clinical relevance threshold for the rescaled efficacy variable. 


For the efficacy curve we observe a maximum absolute distance of approximately $0.0958$ at dose $0.35$, and for toxicity, a maximum distance of $0.0385$ at the highest dose. This results in a maximum of maxima distance of about $0.096$. For toxicity, our findings align closely with those of \citet{Moellenhoff2021}, suggesting that our model is not sensitive to the copula used or the continuous modelling of efficacy.

\begin{figure}[htbp]
    \centering
    \includegraphics[scale = 0.65]{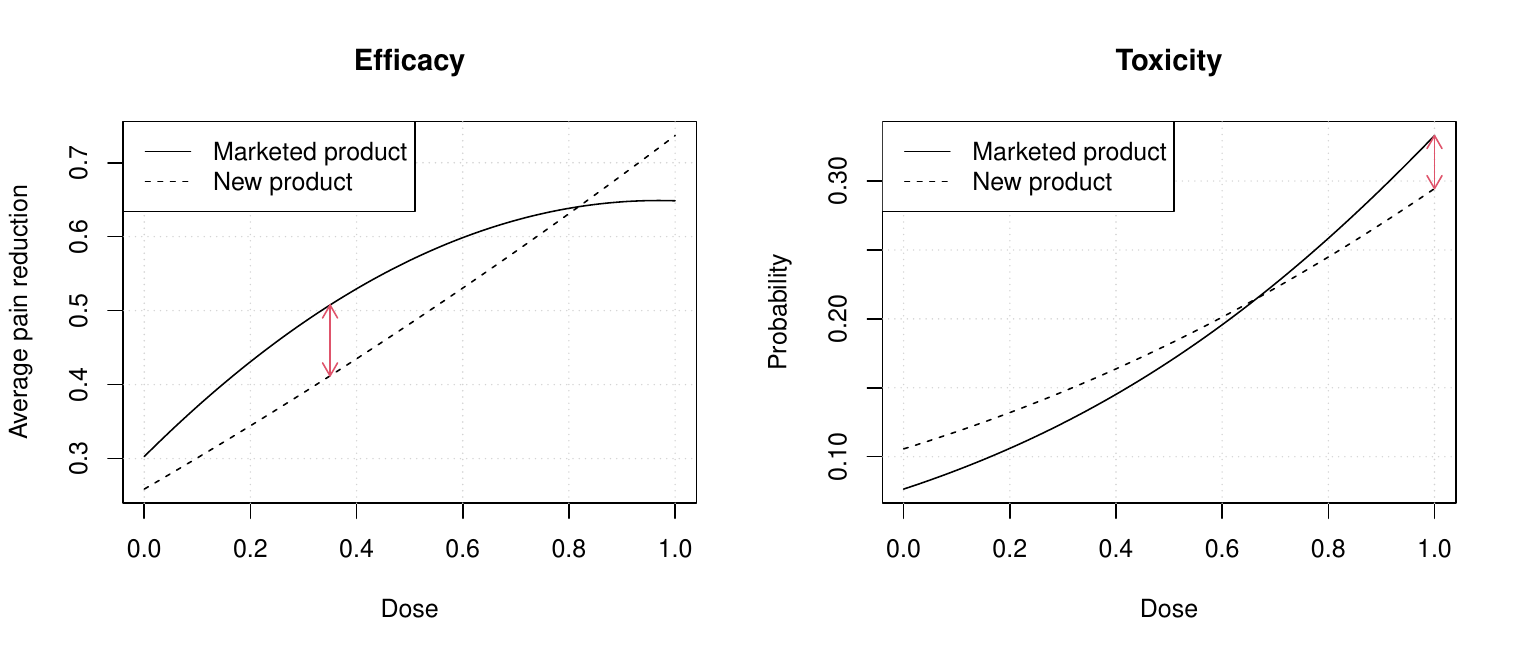}
    \caption{Estimated dose response curves for the efficacy and toxicity of the marketed product and the new product, respectively. The red arrows indicate the maximum distance between the curves.}
    \label{fig:case_study}
\end{figure}

We fit two bivariate mixed outcome models based on the Gaussian copula (as introduced in Section \ref{copula}), one for the new product and the other for the marketed product. We assume a quadratic dose-response curve for the continuous efficacy variable and a logit model for the binary toxicity outcome. Table \ref{tab:case_study} shows the estimated coefficients and Figure \ref{fig:case_study} shows the corresponding dose-response curves. 
For toxicity, we obtain nearly the same coefficients as found by \citet{Moellenhoff2021}. This indicates that neither using a different copula nor modelling the efficacy continuously has a strong influence on the estimates for toxicity. Regarding efficacy, we observe the maximum absolute distance between the curves of approximately $0.0958$ at dose $0.35$. For toxicity, we observe the maximum absolute distance of approximately $0.0385$ for the maximum dose of $1$. Accordingly, the maximum of maxima distance is given by $\hat{d}_{max} = \hat{d}_{\text{Efficacy}} \approx 0.096$. 

\begin{table}[htbp]
  \centering
  \caption{Coefficient estimates for the case study.}
    \begin{tabular}{cccc}
    \hline
          &       & Marketed product & New product   \\
    \hline
    \multirow{3}[0]{*}{Efficacy} & Intercept & 0.303 & 0.259 \\
          & Dose  & 0.715 & 0.416 \\
          & $\text{Dose}^2$ & -0.369 & 0.062 \\
    \multirow{2}[0]{*}{Toxicity} & Intercept & -2.492 & -2.136 \\
          & Dose  & 1.797 & 1.263 \\
    \hline
    \end{tabular}%
  \label{tab:case_study}%
\end{table}%


\begin{table}[htbp]
  \centering
  \caption{Results of the test on similarity proposed in Algorithm \ref{alg1} for different choices of $\epsilon$. The quantiles are obtained using a nominal level of $\alpha=0.05$. Numbers in bold indicate significant p-values at the $5\%$-level.}
    \begin{tabular}{cccc}
    \hline
    $\epsilon$     & $\hat{d}_{max}$     & $\hat{d}^*_{max,(\lfloor n_{boot} \alpha \rfloor )}$  & p-value   \\
    \hline
    0.2   & 0.096 & 0.144 & \textbf{0.003}  \\
    0.15  & 0.096 & 0.105 & \textbf{0.023}  \\
    0.1   & 0.096 & 0.078 & 0.136  \\
    \hline
    \end{tabular}%
  \label{tab:casestudyResult}%
\end{table}%

To assess similarity, we apply Algorithm \ref{alg1} to test the hypotheses \eqref{Hmax}. 
Table \ref{tab:casestudyResult} displays results for varying $\epsilon$ values. For $\epsilon = 0.2, 0.15$, we reject the null hypothesis, suggesting similarity. However, for $\epsilon = 0.1$, we cannot reject H$_0$. These findings mostly align with those of \citet{Moellenhoff2021} for $\epsilon = 0.2$ and $\epsilon = 0.1$.
Unlike \citet{Moellenhoff2021}, our approach allows rejecting H$_0$ even at a more liberal threshold ($\epsilon = 0.15$), indicating the benefit of using a continuous efficacy variable over a binary one. This aligns with our simulation study, which showed higher power for bivariate mixed outcomes as compared to binary ones. The increased power of the proposal is evidenced by the ability to conclude treatment similarity at a 5 \% level for $\epsilon=0.15$, hence highlighting its potential impact in clinical research.

\section{Conclusion}

We introduced a novel model-based equivalence testing approach for multivariate responses, leveraging the flexibility of \emph{generalised joined regression models}. This method stands out for its versatility across various modelling problems, particularly benefiting from the Gaussian copula's capacity to generalise to multi-dimensional settings.
In contrast to existing approaches, our proposal is not limited to univariate or bivariate outcomes allowing for multivariate responses of arbitrary dimension. It accommodates different scales of measures of the outcome variables (e.g., continuous, binary, categorical or ordinal). 
Additionally, we propose an alternative, less conservative testing procedure that contrasts with the intersection union principle. 

The simulation study demonstrates that our method effectively maintains the type 1 error rate at or below the $5 \%$ nominal significance level as sample sizes increase, despite some inflation at smaller sizes for any of the investigated type of outcomes. 
Additionally, we achieve reasonably power values that converge to 1 as sample sizes increase.
Note that we do not observe type I error rates as large as they do, even though we use a less conservative testing procedure. 
For large sample sizes, both approaches achieve reasonable power. However, at small sample sizes, our new approach outperforms the procedure of  \citet{Moellenhoff2021}.


Future possible research includes extending the generalised joint regression models to more than three dimensions, a limitation of the current implementation of the proposed approach. The sensitivity of results to the assumption of Gaussian copula in various contexts, as well as alternative copula options, merits further exploration. Additionally, we aim to adapt the testing procedure for less standard distributions and explore spline-based regression curve specifications.

\section*{Software and data availability}
Software in the form of R code together with the case study data set is available 
via the DOI \href{https://www.doi.org/10.5281/zenodo.10479039}{10.5281/zenodo.10479039}.
\section*{Acknowledgements}

The authors would like to thank Prof. Holger Dette for fruitful discussions during the early phase of developing the method proposed in this paper. 

\section*{Funding}
This work has been supported by the Research Training Group "Biostatistical Methods for High-Dimensional Data in Toxicology`` (RTG 2624, P7) funded by the Deutsche Forschungsgemeinschaft (DFG, German Research Foundation, Project Number 427806116).

\section*{Competing interests}
The authors declare no competing interests.

\end{document}